\documentclass[12pt]{article}
\usepackage{amsfonts}
\usepackage{epsfig}

\newcommand{\Lmat}{L_\mathrm{matter}}

\def\be{\begin{eqnarray}}
\def\ee{\end{eqnarray}}
\def\beann{\begin{eqnarray*}}
\def\eeann{\end{eqnarray*}}
\def\beq{\begin{equation}}
\def\eeq{\end{equation}}
\def\ba{\begin{array}}
\def\ea{\end{array}}
\def\ben{\begin{enumerate}}
\def\een{\end{enumerate}}
\def\bea{\begin{eqnarray}}
\def\eea{\end{eqnarray}}
\def\beann{\begin{eqnarray*}}
\def\eeann{\end{eqnarray*}}
\def\beq{\begin{equation}}
\def\eeq{\end{equation}}
\def\ba{\begin{array}}
\def\ea{\end{array}}
\def\ben{\begin{enumerate}}
\def\een{\end{enumerate}}

\def\5{\bar }
\def\6{\partial }
\def\7{\hat }
\def\4{\tilde }

\def\cH{{\cal H}}

\def\s0#1#2{\mbox{\small{$\frac{#1}{#2}$}}}

\def\qed{\hbox{${\vcenter{\vbox{
\hrule height 0.4pt\hbox{\vrule width 0.4pt height 6pt
\kern5pt\vrule width 0.4pt}\hrule height 0.4pt}}}$}}

\begin{document}
\begin{titlepage}
\begin{flushright}
ULB-TH/02-32\\
hep-th/0211031
\end{flushright}

\begin{centering}

\vspace{0.5cm}

{\bf{\Large Conserved charges in gravitational theories:
  contribution from scalar fields}} \\

\vspace{1.5cm}

{\large Glenn Barnich$^{*}$}

\vspace{.5cm}

Physique Th\'eorique et Math\'ematique,\\ Universit\'e Libre de
Bruxelles,\\
Campus Plaine C.P. 231, B--1050 Bruxelles, Belgium

\vspace{1cm}

\end{centering}
\vspace{.5cm}

\begin{abstract}
In order to illustrate a recently derived covariant formalism for computing
asymptotic symmetries and asymptotically conserved superpotentials in
gauge theories, the case of gravity with minimally coupled scalar fields is
considered and the matter contribution to the
asymptotically conserved superpotentials is computed. 
\end{abstract}

\vfill

\footnotesize{$^*$Research Associate of the Belgium National Fund for
Scientific Research.}

\end{titlepage}

\section{Introduction}
 
Noether currents associated to
gauge symmetries are problematic because they 
are given on-shell by the divergence of 
arbitrary superpotentials. It can however be shown
\cite{Barnich:1995db,Barnich:2000zw}
that reducibility parameters,
i.e., parameters of gauge
symmetries that leave the solutions of the field equations invariant,
are related to conserved superpotentials.
Furthermore, if one defines
suitable equivalence classes of reducibility parameters and of
conserved superpotential, this relation becomes one-to-one and onto, 
and the algebra associated with the
classes of reducibility parameters can be represented in a covariant way
on the classes of the superpotentials \cite{Barnich:2001jy}. 

In full semi-simple Yang-Mills
theory or in full general relativity, there are in fact no non trivial
reducibility parameters and thus also no non trivial conserved
superpotentials. However, non trivial results are obtained if one considers 
the linearized theory around a given background solution. This
observation serves as a starting point for developing a systematic
and covariant theory of asymptotic symmetries and conservation laws
for generic gauge theories
\cite{Anderson:1996sc,Torre:1997cd,Barnich:2001jy}.

Concerning the choice of the boundary and fall-off conditions, 
two approaches are possible. 

The first
is to define from the outset what one means by a field that is
``asymptotically background'' by specifying  
the boundary and the fall-off conditions on these fields. 
The formalism developed in \cite{Barnich:2001jy} 
then allows to define and compute asymptotic reducibility parameters
and the associated asymptotically conserved $n-2$ forms. 

An alternative approach is to first compute the exact reducibility
parameters and the associated conserved $n-2$ forms 
of the linearized theory around the background. For a given
boundary, their asymptotic behaviour can then be determined. 
In this case, the formalism of \cite{Barnich:2001jy} is used not only to 
define asymptotic reducibility
parameters and the associated asymptotically conserved $n-2$ forms,
but also to specify boundary conditions on the left hand sides of the 
field equations. 
These latter conditions then serve to define fields that are
``asymptotically background'': their fall-off is such that, when
inserted in the left hand sides of the field equations, the boundary
conditions are satisfied. 

The way superpotentials are associated to reducibility parameters is
controled by the linearized theory and is independent of the 
choice for the boundary and the fall-off conditions. 
In this short note, we apply the corresponding part of 
the formalism of \cite{Barnich:2001jy}
in the context of matter coupled gravity 
to derive the matter contribution to the
(asymptotically) conserved $n-2$ forms, and thus also to the
charges. Even though various types
of matter fields can be dealt with in a straightforward way, 
we consider here for simplicity only minimally
coupled scalar fields. 

\section{Contribution of scalars to superpotentials}

As in section 6.3 of \cite{Barnich:2001jy}, we suppose that the
spacetime dimension is $n\geq 3$ and that the Lagrangian is given by 
\bea
L=\frac{1}{16\pi}\sqrt{-g}(R-2\Lambda)+\Lmat.
\eea
In the present case, the matter Lagrangian is explicitly given by 
\bea
\Lmat=-\frac{\sqrt{-g}}{8\pi}[g^{\alpha\beta}\6_\alpha\phi^i\6_\beta\phi_i
+V(\phi)].
\eea
The metric is decomposed as $g_{\mu\nu}=\bar g_{\mu\nu}+h_{\mu\nu}$, 
where $\bar g_{\mu\nu}$ denotes the background
metric, which together with its inverse is used to lower and 
raise the indices in the linearized theory, while 
$h_{\mu\nu}$ denotes the metric perturbations. Similarily, 
the scalar fields are decomposed as $\phi^i=\bar\phi^i+\varphi^i$. 
The exact reducibility parameters of the linearized theory are given by
the Killing vectors $\xi^\mu$ 
of the background metric, $L_\xi\bar g_{\mu\nu}=0$,
that satisfy in addition $L_\xi\bar \phi^i=0$, while the asymptotic
reducibility parameters are the vectors $\xi^\mu$  that satisfy 
the asymptotic counterparts (see \cite{Barnich:2001jy}) of these two
conditions.   

The linearized equations of motion are
\bea
\cH^{\mu\nu}
+\frac {\sqrt{-\5g}}{2}\,T^{\mu\nu}_\mathrm{lin}=0,
\eea
where $\cH^{\mu\nu}[h;\5g]$ is the linear
part of $\delta [(1/16\pi)\sqrt{-g}(R-2\Lambda)]/\delta g_{\mu\nu}$,
(see e.g. (6.13) of \cite{Barnich:2001jy}) for the explicit
expression) while 
${\sqrt{-\5g}}\,T^{\mu\nu}_\mathrm{lin}/{2}$ 
is the linear part of 
${\sqrt{-g}}T^{\mu\nu}/{2}={\delta \Lmat}/{\delta g_{\mu\nu}}$. Explicitly,
\bea
\frac{\sqrt{-\5g}}{2}\,T^{\mu\nu}_\mathrm{lin}=\frac{\sqrt{-\5g}}{8\pi}[
\partial^\mu\varphi^i\partial^\nu\bar\phi^i+
\partial^\mu\bar\phi^i\partial^\nu\varphi^i+ \dots,
\eea
where the dots denote terms that do not involve derivatives of
$h_{\mu\nu}$ or $\varphi^i$. 
Because the gauge transformations $L_\xi\bar \phi^i$ do not involve
derivatives of the gauge parameters, the only 
contribution from the scalar fields to the 
$n-1$ form $(d^{n-1}x)_\mu\,
s^\mu_\xi$ (associated to the (asymptotic) reducibility parameters 
$\xi^\mu$ 
and used to
determine the asymptotically conserved $n-2$ form according 
to (1.13) of \cite{Barnich:2001jy}) is given by 
\bea
(d^{n-1}x)_\mu {\sqrt{-\5g}}T^{\mu\rho}_\mathrm{lin}\xi_\rho,
\eea
The contribution from the scalar fields 
to the (asymptotically) conserved $n-2$ form 
$\4k_\xi^{[\nu\mu]}(d^{n-2}x)_{\nu\mu}$ is in turn given by 
\bea
(d^{n-2}x)_{\nu\mu}[\varphi^i\frac{\partial}{\partial \varphi^i_\nu}
{\sqrt{-\5g}}T^{\mu\rho}_\mathrm{lin}\xi_\rho-(\mu\leftrightarrow
\nu)]=\nonumber\\= 
(d^{n-2}x)_{\nu\mu}\frac{\sqrt{-\5g}}{4\pi}\varphi_i[\partial^\mu
\bar\phi^i\xi^\nu-\partial^\nu\bar\phi^i\xi^\mu]. 
\eea
In the three dimensional case with coordinates $t,r,\theta$, 
and boundary the circle at $r\longrightarrow \infty$, 
the matter contribution to the charges $Q_\xi$ is  
\bea
\lim_{r\rightarrow\infty}\int_0^{2\pi}d\theta\,
\frac{\sqrt{-\5g}}{4\pi}\varphi_i[\partial^r
\bar\phi^i\xi^t-\partial^t\bar\phi^i\xi^r].
\eea
The contribution of the scalar fields to these charges has recently
turned out to be relevant \cite{Henneaux:2002wm} in the context of
black hole solutions of $2+1$ gravity with a scalar field. 

\section*{Acknowledgements}

The author wants to thank M.~Henneaux for a useful discussion. 
This work is supported in part by the
``Actions de Recherche Concert\'ees'' of the ``Direction de la Recherche
Scientifique-Communaut\'e Fran\c caise de Belgique, by a ``P\^ole
d'Attraction Interuniversitaire'' (Belgium), by IISN-Belgium
(convention 4.4505.86), by Proyectos FONDECYT 1970151 and 7960001
(Chile) and by the European Commission RTN programme HPRN-CT00131, in
which the author is associated to K.~U.~Leuven.

\vfill
\pagebreak

\end{document}